\documentclass[12pt]{iopart}
\expandafter\let\csname equation*\endcsname=\relax
\expandafter\let\csname endequation*\endcsname=\relax
\usepackage{keyval,xparse}

\newenvironment{frontmatter}{}{}
\makeatletter
\define@key{affiliation}{organization}{\def\affiliation@organization{#1}}
\define@key{affiliation}{addressline}{\def\affiliation@addressline{#1}}
\define@key{affiliation}{postcode}{\def\affiliation@postcode{#1}}
\define@key{affiliation}{city}{\def\affiliation@city{#1}}
\define@key{affiliation}{country}{\def\affiliation@country{#1}}

\DeclareDocumentCommand\affiliation{o m}{%
	\setkeys{affiliation}{organization={},addressline={},postcode={},city={},country={},#2}%
	\address{\affiliation@organization , \affiliation@addressline , \affiliation@postcode\ \affiliation@city, \affiliation@country}%
}
\makeatother	

\usepackage{amssymb}

\usepackage[mathlines]{lineno}

\let\oldequation\equation
\let\oldendequation\endequation
\renewenvironment{equation}
{\linenomathNonumbers\oldequation}
{\oldendequation\endlinenomath}

\usepackage{lmodern}

\usepackage[T1]{fontenc}
\usepackage[utf8]{inputenc}
\usepackage[american]{babel}
\usepackage{color}
\usepackage{float}
\usepackage{textcomp}
\usepackage{mathtools}
\usepackage{amsmath}
\usepackage{amssymb}
\usepackage{amsfonts}
\usepackage{graphicx}
\usepackage{siunitx}
\usepackage{booktabs}

\usepackage{algorithm}
\usepackage{algorithmicx}
\usepackage{algpseudocode}
\algrenewcommand{\algorithmiccomment}[1]{\hfill\textit{*#1*}}
\usepackage{setspace}

\usepackage{caption}
\usepackage{subcaption}
\usepackage{csquotes}

\usepackage[inline]{enumitem}

\usepackage[hidelinks]{hyperref}
\usepackage{cleveref}
\crefname{appendix}{}{}

\usepackage{pgfplots}
\usepackage{tikz}
\pgfplotsset{compat=1.9}

\newlength{\figurewidth}
\newlength{\figureheight}

\tikzset{
	every picture/.append style={
		line join=round,
	}
}
	
\newcommand{\subfolder}{.}

\usetikzlibrary{external}
\usetikzlibrary{positioning,shapes,shadows,arrows.meta,arrows,backgrounds}

\newboolean{forcetikzexternal}
\setboolean{forcetikzexternal}{true}

\NewDocumentCommand{\includetikz}{O{.tikz} O{\subfolder/tikz/} O{\subfolder/tikz_ext/} m}{%
	\ifthenelse{\boolean{forcetikzexternal}}{%
		\includegraphics{#3#4.pdf}%
	}{%
		\tikzsetnextfilename{#3#4}%
		\input{#2#4#1}%
	}%
}

\makeatletter
\let\c@author\relax
\makeatother

\usepackage[natbib=true,%
backend=biber,%
maxcitenames=2,%
maxbibnames=3,%
style=authoryear-comp,%
firstinits=true,%
uniquename=false,%
uniquelist=false,%
bibencoding=utf8,%
doi=true,%
isbn=false,%
url=false,%
eprint=false%
]{biblatex}
\providecommand{\U}[1]{\protect\rule{.1in}{.1in}}
\newtheorem{theorem}{Theorem}%
\newtheorem{problem}[theorem]{Problem}%

\usepackage{acronym}

\newacro{VOI}{volume of interest}
\newacro{OAR}{organ-at-risk}
\newacro{IMRT}{intensity-modulated radiation therapy}
\newacro{MCO}{multi-criteria optimization}
\newacro{SM}{superiorization method}
\newacro{CT}{computed tomography}
\newacro{TV}{total variation}
\newacro{SQP}{sequential quadratic programming}
\newacro{BIP}{block-iterative projection}
\newacro{SAP}{string-averaging projection}
\newacro{AMS}{Agmon, Motzkin, and Schoenberg}
\newacro{DVH}{dose-volume histogram}
\newacro{PTV}{planning target volume}

\usepackage{datetime2}

\DTMsavedate{date_v1}{2022-07-26}

\begin{document}

\begin{frontmatter}

\title[Superiorization for constrained inverse radiotherapy treatment planning]{Superiorization as a novel strategy for linearly constrained inverse radiotherapy treatment planning}

\author[DKFZ,HIRO]{Florian Barkmann}
\ead{flobarkmann@gmail.com}

\affiliation[DKFZ]{organization={Department of Medical Physics in Radiation Oncology, German Cancer Research Center -- DKFZ},
            addressline={Im Neuenheimer Feld 280}, 
            postcode={69120}, 
            city={Heidelberg},
            country={Germany}}

\affiliation[HIRO]{organization={Heidelberg Institute for Radiation Oncology -- HIRO},
	addressline={Im Neuenheimer Feld 280}, 
	postcode={69120}, 
	city={Heidelberg},
	country={Germany}}

\author[HAIFA]{Yair Censor}
\ead{yair@math.haifa.ac.il}

\affiliation[HAIFA]{organization={Department of Mathematics, University of Haifa},
            addressline={Mt.\ Carmel}, 
            city={Haifa},
            postcode={3498838}, 
            country={Israel}}

\author[DKFZ,HIRO]{Niklas Wahl}
\ead{n.wahl@dkfz.de}

\affiliation[DKFZ]{organization={Department of Medical Physics in Radiation Oncology, German Cancer Research Center -- DKFZ},
	addressline={Im Neuenheimer Feld 280}, 
	postcode={69120}, 
	city={Heidelberg},
	country={Germany}}

\affiliation[HIRO]{organization={Heidelberg Institute for Radiation Oncology -- HIRO},addressline={Im Neuenheimer Feld 280}, 
     postcode={69120}, 
     city={Heidelberg},
     country={Germany}}

\begin{abstract}
	\newline
\textbf{Objective:}
We apply the superiorization methodology to the intensity-modulated radiation therapy (IMRT) treatment planning problem. In superiorization, linear voxel dose inequality constraints are the fundamental modeling tool within which a feasibility-seeking projection algorithm will seek a feasible point. This algorithm is then perturbed with gradient descent steps to reduce a nonlinear objective function.

\noindent\textbf{Approach:} 
Within the open-source inverse planning toolkit matRad, we implement a prototypical algorithmic framework for superiorization using the well-established Agmon, Motzkin, and Schoenberg  (AMS) feasibility-seeking projection algorithm and common nonlinear dose optimization objective functions. Based on this prototype, we apply superiorization to intensity-modulated radiation therapy treatment planning and compare its performance with feasibility-seeking and nonlinear constrained optimization. For these comparisons, we use the TG119 water phantom and a head-and-neck patient of the CORT dataset.

\noindent\textbf{Main Results:}
Bare feasibility-seeking with AMS confirms previous studies, showing it can find solutions that are nearly equivalent to those found by the established piece-wise least-squares optimization approach. The superiorization prototype solved the linearly constrained planning problem with similar performance to that of a general-purpose nonlinear constrained optimizer while showing smooth convergence in both constraint proximity and objective function reduction.

\noindent\textbf{Significance:}
Superiorization is a useful alternative to constrained optimization in radiotherapy inverse treatment planning. Future extensions with other approaches to feasibility-seeking, e.\,g., with dose-volume constraints and more sophisticated perturbations, may unlock its full potential for high-performant inverse treatment planning.\\

\noindent Date: \DTMusedate{date_v1}
\end{abstract}

\end{frontmatter}


\section{Introduction}
\label{sect:intro}

Numerical optimization methods lie at the heart of state-of-the-art
inverse treatment planning for \ac{IMRT} \citep{Censor2003,Bortfeld2006a,Censor2012}. Usually, a clinical prescription of the treatment goals forms the input to a nonlinear \ac{MCO} problem with or without additional constraints, depending on the desired patient dose distribution.

During the translation of the clinical goals into an \ac{MCO} problem, one distinguishes between objectives, i.\,e., soft goals that compete with each other, and hard constraints designed to ensure, for example, maximal tolerance doses in an \ac{OAR} and minimal dosage of the target. This versatile approach enables the treatment planner to employ arbitrary combinations of suitable (convex) nonlinear objective functions along with any choice of constraints on the voxels' doses.

This mathematical modeling allows numerical optimization of the fluence of beam elements (beamlets) using a pre-computed normalized dose
mapping \citep{Wu2000}. The resulting constrained nonlinear optimization problem is frequently solved by applying an extended (quasi-)Newton approach with \ac{SQP} or interior-point methods \citep{Bortfeld2006a,Breedveld2017,Fogliata2007,Wieser2017,Wu2000,Luenberger2008,Bazaraa2006}.
Recent work has substantially extended the capabilities of inverse planning through multi-criteria Pareto optimization with subsequent exploration of the Pareto surface or stochastic/robust optimization \citep{Kufer2005,Thieke2007,Unkelbach2018}.

Computational difficulties may arise in the constrained nonlinear optimization approach. Successful optimizers for nonlinear constrained optimization transform the constrained problem into an unconstrained problem using, for example, barrier functions \citep[in the case of interior-point methods, e.\,g.,][]{Wachter2006,Breedveld2017} and the method of Lagrange multipliers in combination with slack variables \citep{Wachter2006,Breedveld2017,Nocedal1999}. This creates a computational burden when the number of constraints increases. Handling many constraints, for example, linear inequalities for some or all individual voxel dose bounds, can inflate the computational effort because each constraint requires a Lagrange multiplier and an additional slack variable. Possible \enquote{workarounds} include minimax-optimization in combination with auxiliary variables or usage of continuous and differentiable maximum approximations like the LogSumExp and softmax functions \citep{Wieser2017}. 

Taking a step back, however, to the starting days of treatment planning research, shows that one does not necessarily need to use an \emph{optimization} approach to solve the purely linearly constrained \ac{IMRT} problem but could use feasibility-seeking projection algorithms \citep{Censor1988,Powlis1989}.

In the context of \ac{IMRT}, such feasibility-seeking translates to seeking a feasible solution that will obey the prescribed lower and upper dose bounds on doses in voxels. If no feasible solution is found, these algorithms find a proximal solution, similar to the piece-wise least-squares approach. Even though they have seen further development over the last decades \citep{Censor1999,Cho2000,Censor2003} and, more recently, also extension to dose-volume constraints \citep{Penfold2017,Brooke2020,Gadoue2022}, numerical optimizers have been the preferred choice in the field due to their abilities to handle the nonlinear objective functions, e.\,g., (generalized) equivalent uniform dose (EUD), which are often desired when prescribing treatment goals. 

The work presented here now combines nonlinear objective functions as used in optimization with feasibility-seeking within linearly constraining dose bounds by applying the \ac{SM}. To do so, the \ac{SM} uses a \emph{superiorized version of the basic algorithm}, the latter being a user-chosen iterative feasibility-seeking algorithm, which is perturbed by interlacing reduction (not necessarily minimization) steps of the chosen (nonlinear) objective function. This practically steers the iterates of the feasibility-seeking algorithm to a feasible solution with a smaller objective function value. 

As a consequence, the \ac{SM} provides a flexible framework for applications and has attracted interest recently: It has demonstrated its effectiveness for image reconstruction in single-energy \ac{CT} \citep{Herman2012,Guenter2022}, dual-energy \ac{CT} \citep{Yang2017} and, more recently, in proton \ac{CT} \citep{Penfold2015,Schultze2020}, by reducing \ac{TV} during image reconstruction. The \ac{SM} has also been successfully applied to diverse other fields of applications, such as tomographic imaging spectrometry \citep{Han2021} or signal recovery \citep{Pakkaranang2020}. For inverse planning, the initial theoretical applicability of superiorization using \ac{TV} as an objective function has been demonstrated in \citet{Davidi2015}.

Building on this preliminary work, we developed, tuned, and evaluated a prototypical superiorization solver for radiotherapy treatment planning problems. To maximize reproducibility and re-usability of our work, our superiorization approach is implemented into the validated open-source radiation therapy treatment planning toolkit matRad \citep{Wieser2017}.

This work is structured as follows: In \cref{sec:methods},
we describe the approaches and present the specific version
of the \ac{SM} that we use along with the feasibility-seeking algorithm
embedded in it. Section \ref{sec:Computational-results} includes
our computational work results. Finally, in \cref{sec:discussion}, we discuss the potential of \ac{SM} with possible future developments and conclude our work in \cref{sec:Conclusions}.

\section{Materials \& Methods}
\label{sec:methods}

This work compares three approaches to model the treatment planning problem in IMRT:
\begin{enumerate*}[label=(\roman*)]
\item the \emph{nonlinear constrained minimization approach} of minimizing an objective function subject to constraints,
\item the \emph{feasibility-seeking approach}
searching for a feasible solution adhering to constraints without
considering any objective functions to minimize, and finally, \item the \emph{superiorization approach}, which perturbs the feasibility-seeking algorithm to reduce (not necessarily minimize) an objective function.
\end{enumerate*}
Before introducing these approaches, we briefly recap the discretization of the inverse treatment planning problem.

\subsection{Discretization of the inverse treatment planning problem}
\label{subsec:discretization}
Computerized inverse treatment planning usually relies on a spatial discretization of the particle fluence, the patient anatomy, and, consequently, the radiation dose. 

The patient is represented by a three-dimensional voxelized grid (image) with $n$ voxels numbered $i = 1,2,\ldots,n$. Based on this image, $Q$ \acp{VOI} $S_q,\: q = 1, 2, \ldots, Q$ are segmented. This allows us to represent the dose as a vector $\boldsymbol{d} = (d_i)^n_{i=1}$, whose $i$-th component is the radiation dose deposited within the $i$-th voxel. For each of the segmentations $S_q$, we can then easily identify its dosage by finding $d_i$ for all $i \in S_q$.

The radiation fluence is represented as a vector intensities $\boldsymbol{x}=(x_{j})_{j=1}^{m}$, whose $j$-th component is the intensity of the $j$-th beamlet. The dose deposition $a_i^j$ for a unit intensity of beamlet $j$ to voxel~$i$ can then be precomputed and stored in the \emph{dose influence matrix} $A = (a_i^j)_{i=1,j=1}^{n,m}$, mapping $\boldsymbol{x}$ to $\boldsymbol{d}$ via $\boldsymbol{d} = A \boldsymbol{x}$.

\subsection{The constrained minimization approach}
\label{subsec:optimization}

In the optimization approach to \ac{IMRT} treatment planning, the clinically prescribed aims are represented by various (commonly differentiable) objective functions which map the vector of beamlet intensities to the positive real numbers \cite{Wu2000}. 

For our purposes, we limit ourselves to objective functions $f_{p}\colon\mathbb{R}^{n}\rightarrow[0,\infty),\;p=1,2,\ldots,P$, operating on the radiation dose $\boldsymbol{d}$ as surrogates for clinical, dose-based goals. 

A comprehensive, exemplary list of such common objective functions can be found in \citet[Table 1]{Wieser2017} and, for the reader's convenience, also in \cref{sec:app_objectives} below. These objective functions, which depend on the dose, are related to the intensities $\boldsymbol{x}$ via $\boldsymbol{d} = A \boldsymbol{x}$, which is computed at each iterate/change of $\boldsymbol{x}$ during optimization.

Wishing to fulfill or decide between multiple clinical goals, the resulting multi-objective optimization problem may be scalarized using a weighted sum of several different individual objective functions for the various \acp{VOI} $S_q$.  This approach, first introduced for least-squares \citep[as introduced by][]{Bortfeld1990}, can today explore a plethora of objective functions \citep{Wu2000,Wieser2017} while also satisfying hard constraints \citep{Wieser2017,Breedveld2017}:

\begin{equation}
\begin{aligned}
\boldsymbol{x}^* & =  &&\underset{\boldsymbol{x}}{\arg\min} \sum_{p=1}^{P}w_{p}f_{p}(\boldsymbol{d}(\boldsymbol{x})) &&& \\
\mathrm{such\ that} & && c^L_t \leq c_t(\boldsymbol{d}(\boldsymbol{x})) \leq c^U_t &&& t = 1, 2, \ldots, T,\\
& &&\boldsymbol{x} \geq 0
\end{aligned}
\label{eq:hybrid-problem}
\end{equation}

Here $w_{p}\geq0$, for all $p=1,2,\ldots,P$, are user-specified weights reflecting relative importance, $f_{p}$ are user-chosen individual objective functions, $\boldsymbol{x}$ is the beamlet radiation intensities vector (which is physically bound to the nonnegative real orthant), and $c_t$ are user-chosen individual constraints with lower and upper bounds $c^L_t$ and $c^U_t$, respectively. While the constraints $c_t$ can, in principle, be nonlinear constraints, we focus here on \emph{linear inequality constraints} representing upper and lower dose prescription bounds.

The inverse planning problem (\ref{eq:hybrid-problem}), solved with numerical optimization techniques, is commonly used today across treatment modalities \citep[among others][]{Bortfeld1990,Wu2000,Alber2007,Wieser2017,Breedveld2017}. \Ac{SQP} or interior-point methods with a (quasi-)Newton approach are often used to solve the resulting constrained optimization problems \citep{Bortfeld2006a,Breedveld2017,Fogliata2007,Wieser2017,Wu2000,Carlsson2006,Bazaraa2006,Luenberger2008}.

\subsection{The feasibility-seeking approach}
\label{subsec:The-pure-feasibility-seeking}

Our work suggests to not formulate the inverse planning as a constrained
optimization problem but as a feasibility-seeking problem for a given
set of constraints. The feasibility-seeking approach has been suggested before in literature \citep[see, e.\,g.,][and references therein]{Censor2006}.

To solve the treatment planning problem with bare feasibility-seeking, we model our dose prescriptions as a system of linear inequalities, which leads to a full discretization of the problem. In general, for every voxel, we define a lower and upper bound of dose and seek
a solution, i.\,e., an intensities vector that fulfills these prescriptions.

The bare feasibility-seeking approach is thus limited to treatment planning problems formulated as a system of linear inequalities and is thereby not as flexible as the constrained optimization approach. Yet, since its formulation is the backbone of the \ac{SM} allowing to extend feasibility-seeking with soft goals, it will be outlined below using the notation from \cref{subsec:discretization,subsec:optimization}.

With $\boldsymbol{d}(\boldsymbol{x})=A\boldsymbol{x}$, the beamlet
radiation intensities vector $\boldsymbol{x}$ now has to be recovered
from a system of linear inequalities of the form 
\begin{equation}
c_{i}^{L}\leq\sum_{j=1}^{m}a_{i}^{j}x_{j}\leq c_{i}^{U},\quad i=1,2,\ldots,n\,.\label{eq:linear_inequalities}
\end{equation}
In principle, individual lower and upper bounds $c_{i}^{L}$ and $c_{i}^{U}$
can be chosen for each voxel~$i$. Since prescriptions are usually
grouped per \ac{VOI} $S_{q}$, the system can be rewritten as:
\begin{equation}
\textup{For all}\ q=1,2,\ldots,Q: \ell_{q}\leq\sum_{j=1}^{m}a_{i}^{j}x_{j}^{*}\leq u_{q}\text{\ for all\ }i\in S_{q},\label{eq:linear_inequalities_voi}
\end{equation}
with $\ell_{q}$ and $u_{q}$ representing the lower and upper dose
bounds per \ac{VOI} $S_{q}$, respectively. Since it does not make
sense to prescribe positive lower bounds to \acp{OAR}, these are
generally chosen to be equal to zero.

Geometrically, depending on which structure $S_{q}$ a voxel $i$
belongs to, each physical dose constraint set $C_{i}$ in each voxel
$i=1,2,\ldots,n,$ is a \emph{hyperslab} (i.e., an intersection of
two half-spaces) in the $m$-dimensional Euclidean vector space $\mathbb{R}^{m}$.

Aiming at satisfaction of all physical dose constraints along with
the nonnegativity constraints is, thus, the following \emph{linear
feasibility problem} \citep[which is a special case of the \emph{convex
feasibility problem}, see, e.\,g.,][]{Bauschke1996,Censor2018}:

\begin{equation}
	\begin{split}
\text{Find an}\ \boldsymbol{x}^{*}\in W\coloneqq\{\boldsymbol{x}\in\mathbb{R}^{m}\,|\text{ for all, } q=1,2,\ldots,Q,\ \ell_{q}\leq\sum_{j=1}^{m}a_{i}^{j}x_{j}\leq u_{q},\\
\text{\ for all\ }i\in S_{q},\textup{ and }\boldsymbol{x}\geq 0\}\label{eq:linear-feas-prob}
\end{split}
\end{equation}

Such feasibility-seeking problems can typically be solved by a variety of efficient projection methods, whose main advantage, which makes them successful in real-world applications, is computational \citep[see, e.g.,][]{Censor2012}.

They commonly can handle very large-size problems of dimensions beyond which other, more sophisticated currently available, methods start to stutter or cease to be efficient. 
This is because the building blocks of a projection algorithm are the projections onto the given individual sets. These projections are actually easy to perform, particularly in linear cases such as hyperplanes, half-spaces, or hyperslabs. 

For the purpose of this paper, we define such an iterative feasibility-seeking algorithm via an algorithmic operator $\mathcal{A}: \mathbb{R}^m \rightarrow \mathbb{R}^m$,
\begin{equation}
\boldsymbol{x}^{0}\in \mathbb{R}^m,\ \boldsymbol{x}^{k+1}=\mathcal{A}(\boldsymbol{x}^{k}),\ k=1,2,\ldots\ ,\label{eq:target-alg}
\end{equation}
whose task is to (asymptotically) find a point in $W$.

The algorithmic structures of projection algorithms are sequential, simultaneous, or in-between, such as in the \ac{BIP} methods \citep[see, e.\,g.,][and references therein]{Davidi2009,Gordon2005} or in the more recent \ac{SAP} methods \citep[see, e.\,g.,][and references therein]{Censor2014a,Bargetz2018}. An advantage of projection methods is that they work with the initial, raw data and do not require transformation of, or other operations on, the sets describing the problem.

For our prototype used here in conjunction with the \ac{SM} described in the sequel, we rely on the well-established \ac{AMS} relaxation method for linear inequalities \citep{Agmon1954,Motzkin1954}. Implemented sequentially and modified for handling the bounds $\boldsymbol{x} \geq 0$, it is outlined in \cref{alg:Sequential-Relaxation-Method}. We denote $\ell:=(\ell_{q})_{q=1}^{Q}$ and $u:=(u_{q})_{q=1}^{Q}$.

\begin{algorithm}
\caption{The \ac{AMS} Sequential Relaxation Method's algorithmic operator $\mathcal{A}^{\text{AMS}}$}
\label{alg:Sequential-Relaxation-Method}
\begin{algorithmic}[1]\onehalfspacing
\Function{$\mathcal{A}^{\text{AMS}}$}{$\boldsymbol{x}, A,u,\ell,\lambda,\nu$}
\State $I = \operatorname{CS}(n)$ \Comment{select control sequence}
\ForAll {$i\in I$}
    \If {$\langle \boldsymbol{a}^{i},\boldsymbol{x}\rangle>u_{q}$} \Comment{$u_{q}$ for structure containing $i$-th voxel}
        \State $\boldsymbol{x}\gets \boldsymbol{x}-\lambda \nu_{i}\dfrac{\langle \boldsymbol{a}^{i},\boldsymbol{x}\rangle-u_{q}}{||\boldsymbol{a}^{i}||_{2}^{2}}\boldsymbol{a}^{i}$
    \ElsIf {$\langle \boldsymbol{a}^{i},\boldsymbol{x}\rangle<\ell_{q}$} \Comment{$\ell_{q}$ for structure containing $i$-th voxel}
    \State $\boldsymbol{x}\gets \boldsymbol{x}-\lambda \nu_{i}\dfrac{\ell_{q} - \langle \boldsymbol{a}^{i},\boldsymbol{x}\rangle}{||\boldsymbol{a}^{i}||_{2}^{2}}\boldsymbol{a}^{i}$
    
    \Else \Comment{Do nothing}
    \EndIf
\EndFor
\For{$j = 1,2,\ldots m$} \Comment{Ensure nonnegativity of $\boldsymbol{x}$}
    \State $x_j \gets \begin{cases} x_j, & \text{for}\ x_j \geq 0\\ 0, & \text{for}\ x_j < 0\end{cases}$
    \EndFor
    \State\Return $\boldsymbol{x}$
\EndFunction
\end{algorithmic}
\end{algorithm}

During an iteration, \cref{alg:Sequential-Relaxation-Method} iterates over all rows of the dose matrix $A$ and handles sequentially the right-hand side and the left-hand side of individual constraints from \cref{eq:linear_inequalities_voi}. The \emph{control sequence} (CS) \citep[Definition 5.1.1]{Censor1998} determines the order of iterating through the matrix rows/constraints. When a corresponding voxel dose inequality is violated, the algorithm performs geometrically a projection of the current point $\boldsymbol{x}$ onto the violated half-space with a user-chosen relaxation parameter $0<\lambda\leq 2$. The original \ac{AMS} algorithm is modified in \cref{alg:Sequential-Relaxation-Method} to allow the relaxation for each voxel~$i$ to be weighted with $\nu_i$ and by performing projections onto the nonnegative orthant of $\mathbb{R}^m$ (in steps 11--13) to return only nonnegative intensities $\boldsymbol{x}$. The vector $\boldsymbol{a}^{i}=(a_{i}^{j})_{j=1}^{m}$ is the $i$-th row of the dose matrix A and is the normal vector to the half-space represented by that row and $||\boldsymbol{a}^{i}||_{2}^{2}$ is its square Euclidean norm.

\subsection{The superiorization method and algorithm}\label{sec:sm}
The \ac{SM} is built upon application of a feasibility-seeking approach (\cref{subsec:The-pure-feasibility-seeking}) to the constraints in the second and third lines of \cref{eq:hybrid-problem}. But instead of handling the constrained minimization problem of \cref{eq:hybrid-problem} with a full-fledged algorithm for constrained minimization, the \ac{SM} interlaces into the feasibility-seeking iterative process (i.\,e., the \enquote{the basic algorithm}) steps that reduce locally in each iteration the objective function value. 

Accordingly, the \ac{SM} does not aim at finding a constrained minimum of the combined objective function $f(\boldsymbol{x})=\sum_{p=1}^{P}w_{p}f_{p}(\boldsymbol{x})$ of \cref{eq:hybrid-problem} over the constraints. It rather strives to find a feasible point that satisfies the constraints and has a reduced -- not necessarily minimal -- value of $f$. 

In the following, we give a brief and focused introduction to \ac{SM}.  A more detailed explanation and review can be found in, e.g., \citet[Section II]{Censor2021} and references therein \citep[see also][]{Censor2015,Censor2014,Censor2015,Davidi2009,Herman2014,Schultze2020,Censor2017,Herman2012}. 

In general, the \ac{SM} is intended for \emph{constrained function reduction problems} of the following form \citep[Problem 1]{Censor2017}:
\begin{problem}[The constrained function reduction problem of the \ac{SM}]\label{prob:sm}$\quad$\newline 
Let $W$ be a given set (such as in \cref{eq:linear-feas-prob}) and let $f:\mathbb{R}^{m}\rightarrow\mathbb{R}$ be an objective function (such as in \cref{eq:hybrid-problem}). Let $\mathcal{A}$ from \cref{eq:target-alg} be an algorithmic operator that defines an iterative \enquote{basic algorithm} for feasibility-seeking of a point in $W$. Find a vector $\boldsymbol{x}^{*}\in W$ whose function value is lesser (but not necessarily minimal) than that of a point in $W$ that would have been reached by applying the basic algorithm alone.
\end{problem}

The \ac{SM} approaches this question by investigating the \emph{perturbation
resilience} \citep[Definitions 4 and 9]{Censor2015} of $\mathcal{A}$, and then proactively using  such perturbations, to locally reduce the values $f$ of the iterates, in order to steer the iterative sequence generated by algorithm $\mathcal{A}$ to a solution with smaller objective function value. The structure of the superiorization algorithm implemented here is given by \cref{alg:superiorization} with explanations here and in \cref{subsec:mod-super}. 

\begin{algorithm}
\caption{The superiorized version of the feasibility-seeking basic algorithm  $\mathcal{A} = \mathcal{A}^{\text{AMS}}$\label{alg:superiorization}}\label{alg_super}

\begin{algorithmic}[1]
    \State $k\gets 0$
    \State $\boldsymbol{x}^{k}\gets \boldsymbol{x}^0$
    \State $s\gets -1$
    \While {stopping rule not met}
        \State $t\gets 0$ \Comment{start of perturbation phase}
        \State $\boldsymbol{x}^{k,t}\gets \boldsymbol{x}^{k}$
        \While {$t < N$} \Comment{apply $N$ function reductions}
            \State $loop\gets \texttt{true}$
            \While{$loop$} 
                \State $s \gets s+1$
                \State $\beta\gets\alpha^s$ \Comment{Step size adaptation}
                \State $\boldsymbol{z}\gets \boldsymbol{x}^{k,t}-\beta\nabla f(\boldsymbol{x}^{k,t})$ \Comment{Function reduction step}
                \If {$f(\boldsymbol{z})\leq f(\boldsymbol{x}^{k,t})$} \Comment{Function reduction check}
                    \State {$t \gets t + 1$}
                    \State $\boldsymbol{x}^{k,t}\gets \boldsymbol{z}$
                    \State $loop\gets \texttt{false}$
                \EndIf
            \EndWhile
        \EndWhile
        \State {$\nu_{k} \gets \eta^k \nu$} \Comment{start of feasibility-seeking phase}
        \State {$\boldsymbol{x}^{k+1}\gets\mathcal{A}^{\text{AMS}}(\boldsymbol{x}^{k,t},A,u,\ell,\lambda,\nu_{k})$}
        \State {$k\gets k+1$}
    \EndWhile
    \State \Return $\boldsymbol{x}^k$
\end{algorithmic}
\end{algorithm}

Except for the initialization in steps 1--3, \cref{alg_super}
consists of the perturbations phase (steps 5--19) and
the feasibility-seeking phase (steps 20--23).

In the perturbation phase, the objective function $f$ is reduced using gradient descent steps. The step-size $\beta$ of these gradient updates is calculated by $\alpha^{s}$ where $\alpha$ is a fixed user-chosen constant, called \emph{kernel}, $0 < \alpha < 1$ so that the resulting step-sizes are nonnegative and form a summable series. The power $s$ is incremented by one until the objective function value of the newly acquired point is smaller or equal to the objective function value of the point with which the current perturbations phase was started. 

The parameter $N$ determines how many perturbations are executed
before applying the next full sweep of the feasibility-seeking phase. The basic \cref{alg:Sequential-Relaxation-Method}  with algorithmic operator $\mathcal{A}^{\text{AMS}}$, used throughout this work,  is indeed \enquote{perturbation resilient} \citep{Censor2013}. 

The superiorization approach has the advantage of letting the user choose any task-specific algorithmic operator $\mathcal{A}$ that will be computationally efficient, independently of the perturbation phase, as long as perturbation resilience is preserved. 

For our \ac{IMRT} treatment planning problem using voxel dose constraints as introduced in \crefrange{eq:linear_inequalities}{eq:linear-feas-prob}, $\mathcal{A}$ can be -- besides the chosen \ac{AMS} algorithm -- any of the wide variety of feasibility-seeking algorithms \citep[see, e.\,g.,][]{Bauschke2013,Cegielski2012,Censor2015a,Censor2012a,Censor1998,Bauschke1996}.

The principles of the \ac{SM} have
been presented and studied in previous publications \citep[consult, e.\,g.][]{Herman2012,Herman2014,Censor2015}, but, to the best of our knowledge, this is the first work applying the \ac{SM} to a treatment planning problem with an objective function of the general form $f(\boldsymbol{x}):=\sum_{p=1}^{P}w_{p}f_{p}(\boldsymbol{x})$ from \cref{eq:hybrid-problem}.

\subsubsection{Modifications of the prototypical superiorization algorithm}
\label{subsec:mod-super}

To control the initial step-size, we \enquote{warm start} the algorithm with larger kernel powers $s$ within the first iteration, which substantially improves the algorithm's run-time. For our purposes, we chose an initial increment of $s \leftarrow s + 25$.

In the feasibility-seeking phase, instead of weighting all projections onto the half-spaces equally via the relaxation parameters, each projection can also be given an individual weight $0<\nu_{i}<1$ representing the \enquote{importance} of the $i$-th inequality constraint (i.\,e., voxel). 

Further, as shown in step 20 of \cref{alg:superiorization}, weights can be reduced after each iteration to improve stability. Similar to how the step-sizes are reduced in the perturbation phase, we utilize another kernel $0<\eta<1$ and use its powers $\eta^{k}$ to reduce the weights in step 20 by incrementing $k$ after each feasibility-seeking sweep. The new weights are then calculated by $\eta^{k}\cdot \nu,$ where $\nu$ are the initial weights.

Finally, we integrate four different control sequences to iterate through the rows of $A$. Apart from following the cyclic order according to voxel indices, we experimented with a random order and with sequences choosing rows with increasing or decreasing weights $\nu_{i}$. 

\subsubsection{Stopping criteria}\label{subsec:Stopping}
The algorithm was terminated after a given maximal number of iterations was reached or after a certain time limit was exceeded, or when the stopping criterion formulated below was met. The default number of maximum iterations was \num{500} and the default wall-clock duration was set to \SI{50}{\minute}. 

The stopping criterion that we used consists of two parts, both of which must be met for three consecutive iterations for the algorithm to stop. The first part of the stopping criterion is that the relative change of the objective function $f$ defined by
\begin{equation}
\left|\dfrac{f(\boldsymbol{x}^{k+1})-f(\boldsymbol{x}^{k})|}{\max\lbrace1,f(\boldsymbol{x}^{k})\rbrace}\right|
\end{equation}
becomes smaller than $10^{-4}$. 

For the second part of the stopping criterion, we define the square of the
weighted $L_{2}$-norm of the constraints violations by\footnote{For any real number $r$ we use: $(r)_{+}:=\max\left\{ 0,r\right\}$.}
\begin{equation}
V(\boldsymbol{x})\coloneqq\dfrac{1}{n}\sum_{i=1}^{n}\dfrac{(\ell_{q}-\langle \boldsymbol{a}^{i},\boldsymbol{x}\rangle)_{+}^{2}+(\langle \boldsymbol{a}^{i},\boldsymbol{x}\rangle-u_{q})_{+}^{2}}{||\boldsymbol{a}^{i}||_{2}^{2}},
\end{equation}
where $\ell_{q}$ and $u_{q}$ depend on which structure the $i$-th
voxel belongs to. This second part of the stopping rule is met if
the relative change of $V$ defined by
\begin{equation}
\left|\dfrac{V(\boldsymbol{x}^{k+1})-V(\boldsymbol{x}^{k})|}{\max\lbrace1,V(\boldsymbol{x}^{k})\rbrace}\right|
\end{equation}
is smaller than $10^{-3}.$ 

All tolerances of the stopping criteria
can be customized and also set to a negative number to turn off single
stopping criteria or early stopping altogether.

\subsection{Implementation}
\label{sec:matRad}
The superiorization prototype described above was implemented in the open-source cross-platform \enquote{matRad} \citep{matRadBlaise,matRad2015,Wieser2017}, which is a multi-modality radiation dose calculation and treatment planning toolkit written in Matlab. The implementation is publicly available on the matRad GitHub repository on a research branch.\footnote{\url{https://github.com/e0404/matRad/tree/research/superiorization}}

The implementation in matRad facilitates comparison against plans generated on the same datasets with a nonlinear optimizer, as matRad implements a number of common objective functions used in treatment planning (compare to \cref{app:objectives} and \citet[Table I]{Wieser2017}). While matRad provides interfaces to both the open-source Interior Point OPTimizer (IPOPT) \citep{Wachter2006} as well as to Matlab's built-in interior-point algorithm from \texttt{fmincon}, only the first was used for our comparisons.

matRad performs all computations in a fully-discretized model with a voxel grid. The \enquote{dose matrix} $A$ is stored as a compressed sparse column matrix computed for all analyses using matRad's singular value decomposed pencil-beam algorithm \citep{Bortfeld1993} validated against a clinical implementation \citep{Wieser2017}. 

\section{Results}
\label{sec:Computational-results}
\subsection{Proof-of-work: Phantom plan}
To demonstrate the applicability of superiorization to the IMRT treatment
planning problem, we first evaluate a small example using the horseshoe
phantom defined in the AAPM TG119 Report \citep{Ezzell2009}. The
phantom is part of the CORT dataset \citep{Craft2014} and consequently
available with matRad. 

We created an equidistantly spaced 5-field \ac{IMRT} photon plan with $\SI{5}{\milli\meter} \times \SI{5}{\milli\meter}$ beamlet doses (resulting in 1918 pencil-beams and a corresponding sparse dose influence
matrix with $9.3\times10^{7}$ non-zero entries in $3.5\times10^{6}$
voxels). 

With this setup, we generated treatment plans using three different approaches: 
\begin{enumerate*}[label=(\roman*)]
    \item constrained minimization with IPOPT,
    \item the \ac{AMS} algorithm for feasibility-seeking only, and 
    \item the \ac{SM} with the \ac{AMS} algorithm.
\end{enumerate*}
Different combinations of nonlinear objective functions and linear inequality constraints on dose were evaluated and compared across these approaches.

For analysis, we use \acp{DVH} and axial dose (difference) slices, as well as the evolution plots of the objective function values and the constraint violations.

\subsubsection{General usability of the \protect\ac{AMS} feasibility-seeking projection algorithm}
\label{subsec:ams_poc}

We first validate that our implemented projection algorithm \ac{AMS} is capable of finding comparable treatment plans to those found by established optimization algorithms when applied to
a straightforward piece-wise least-squares objective function for the unconstrained minimization of residuals. 

The setup prescribes \SI{60}{\gray} to the C-shaped target. To achieve this prescription, we bound the dose in the target by \SI[separate-uncertainty = true]{60+-1}{\gray}. To the two \acp{OAR}, \enquote{Core} and \enquote{Body}, upper bounds (a.\,k.\,a.\ tolerance doses) are prescribed, resulting in the parameters given in \cref{tab:optams1}. 

For nonlinear minimization with IPOPT, the tolerance doses serve as parameters for respective penalized piece-wise least-squares objective functions while for \ac{AMS} the tolerances directly translate into linear inequalities and the weights proportionally increase the relaxation parameters.

\begin{table}[htb]
\centering
\caption{Dose inequalities / prescriptions and penalty weights used for minimization and for \ac{AMS} feasibility-seeking.}
\label{tab:optams1}
\small
\begin{tabular}{ccc}
\toprule 
VOI & $w_{p}$ & tolerance / inequality constraint\tabularnewline
\midrule
Target & 1000 & $\SI{59}{\gray} < \boldsymbol{d} < \SI{61}{\gray}$ \tabularnewline
Core & 100 & $\boldsymbol{d} < \SI{20}{\gray}$ \tabularnewline
Body & 30 & $\boldsymbol{d} < \SI{30}{\gray}$ \tabularnewline
\bottomrule
\end{tabular}
\end{table}

\begin{figure}[bth]
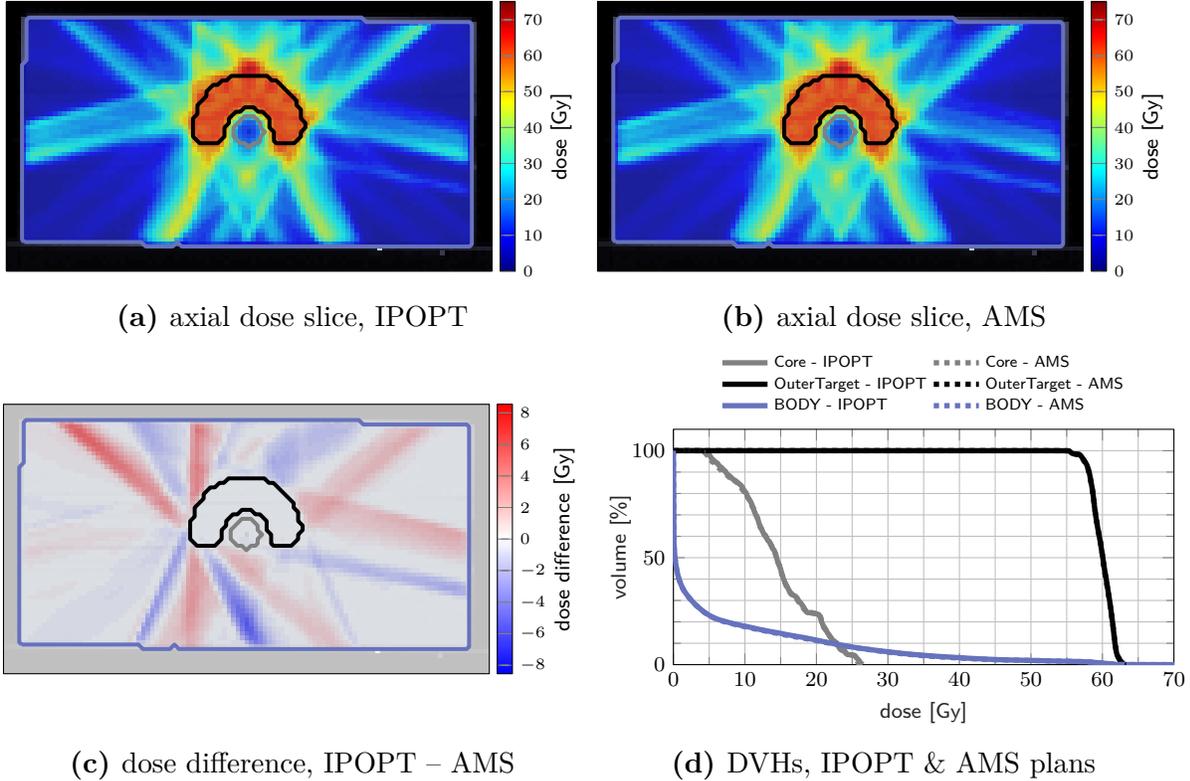

    \begin{subfigure}[b]{0.499\textwidth}
    	\centering%
    	\scriptsize\sffamily%
    	\setlength{\figurewidth}{0.825\textwidth}%
    	\setlength{\figureheight}{0.555\figurewidth}%
        \includetikz{super_script_basic_ipoptPlan}%
        \caption{axial dose slice, IPOPT}
    \end{subfigure}%
    \hfill%
    \begin{subfigure}[b]{0.499\textwidth}
    	\centering%
    	\scriptsize\sffamily%
    	\setlength{\figurewidth}{0.825\textwidth}%
    	\setlength{\figureheight}{0.555\figurewidth}%
    	\includetikz{super_script_basic_amsPlan}%
        \caption{axial dose slice, \ac{AMS}}
    \end{subfigure}%
	\vspace{1mm}%
    
    \begin{subfigure}[b]{0.499\textwidth}
       	\centering%
    	\scriptsize\sffamily%
    	\setlength{\figurewidth}{0.825\textwidth}%
    	\setlength{\figureheight}{0.555\figurewidth}%
    	\includetikz{super_script_basic_diffPlan}%
      	\captionsetup{skip=3em}%
        \caption{dose difference, IPOPT -- \ac{AMS}}
    \end{subfigure}%
    \hfill%
    \begin{subfigure}[b]{0.499\textwidth}
       	\centering%
    	\scriptsize\sffamily%
    	\setlength{\figurewidth}{0.85\textwidth}%
    	\setlength{\figureheight}{0.4\textwidth}%
        \includetikz{super_script_basic_dvhs}%
        \caption{\acp{DVH}, IPOPT \& \ac{AMS} plans}
    \end{subfigure}
    \caption{Comparison of treatment plans obtained by nonlinear minimization with IPOPT and by feasibility-seeking with \ac{AMS}, using the tolerances from \cref{tab:optams1}.}
    \label{fig:AMS_poc}
\end{figure}

\Cref{fig:AMS_poc} confirms that feasibility-seeking with
weighted \ac{AMS} is able to find dose distributions of similar quality
as conventional nonlinear unconstrained minimization of a piece-wise least-squares objective function. While resulting in different intensity-modulation patterns, nearly congruent \acp{DVH} are observed. 

A crude performance analysis though measures substantially longer run-times for the \ac{AMS} approach (about five times slower than unconstrained minimization). This difference is mainly driven by the fact that \ac{AMS} does sequential iteration through the matrix rows in each sweep.

This investigated scenario is, however, not intended to display any performance advantages of the \ac{AMS} algorithm, but only to validate  its behavior and confirm the long-known ability of such feasibility-seeking algorithms to yield acceptable treatment plans \citep{Censor1988,Powlis1989}.

\subsubsection{Inverse planning with superiorization}
\label{sec:super_planning}
Using the same phantom and irradiation geometry as in section \cref{subsec:ams_poc}, the feasibility problem used in \ref{subsec:ams_poc} was modified to enforce some hard
linear inequality constraints while minimizing an objective function. When the constraints are feasible, superiorization using \ac{AMS} as the basic algorithm will find a feasible point while perturbing the iterates of the feasibility-seeking algorithm towards smaller (not necessarily minimal) function values with objective function reduction steps.

As reference, nonlinear constrained minimization with IPOPT with a logistic maximum approximation for minimum/maximum \citep[compare][Table 1]{Wieser2017}, was used. Three prescription scenarios were investigated:
\begin{enumerate*}[label=(\Roman*)]
    \item linear inequalities on the target $(\SI{59}{\gray} < \boldsymbol{d} < \SI{61}{\gray})$, 
    \item additional linear inequalities on the \enquote{Core} structure $(\boldsymbol{d} < \SI{30}{\gray})$, and 
    \item only linear inequalities on the \enquote{Core} $(\boldsymbol{d} < \SI{30}{\gray})$. 
\end{enumerate*}
The parameters are detailed in \cref{tab:super_poc_settings}.

\begin{table}[htbp]
\caption{Dose inequality constraints, objective functions, and penalty weights used separately for constrained minimization and for superiorization. The Roman numerals in parentheses
for the inequality constraints describe their usage in the plans, respectively. The functions in the right-hand column stem from \citet[Table 1]{Wieser2017} and are identified here in \cref{app:objectives} below.}
\label{tab:super_poc_settings}
\small
\centering
\begin{tabular}{cccc}
\toprule
VOI & $w_p$ & $c(\boldsymbol{d})$ & $f(\boldsymbol{d})$\tabularnewline
\midrule
Target & 1000 & $\SI{59}{\gray} < \boldsymbol{d} < \SI{61}{\gray}$ (I \& II) & $f_{\mathrm{sqdev}}(\boldsymbol{d};\SI{60}{\gray})$\tabularnewline
Core & 100 & $\boldsymbol{d} < \SI{30}{\gray}$ (II \& III) & $f_{\mathrm{sqdev}+}(\boldsymbol{d};\SI{20}{\gray})$\tabularnewline
Body & 30 & - & $f_{\mathrm{sqdev}+}(\boldsymbol{d};\SI{30}{\gray})$\tabularnewline
\bottomrule
\end{tabular}
\end{table}

\begin{figure}[htbp]
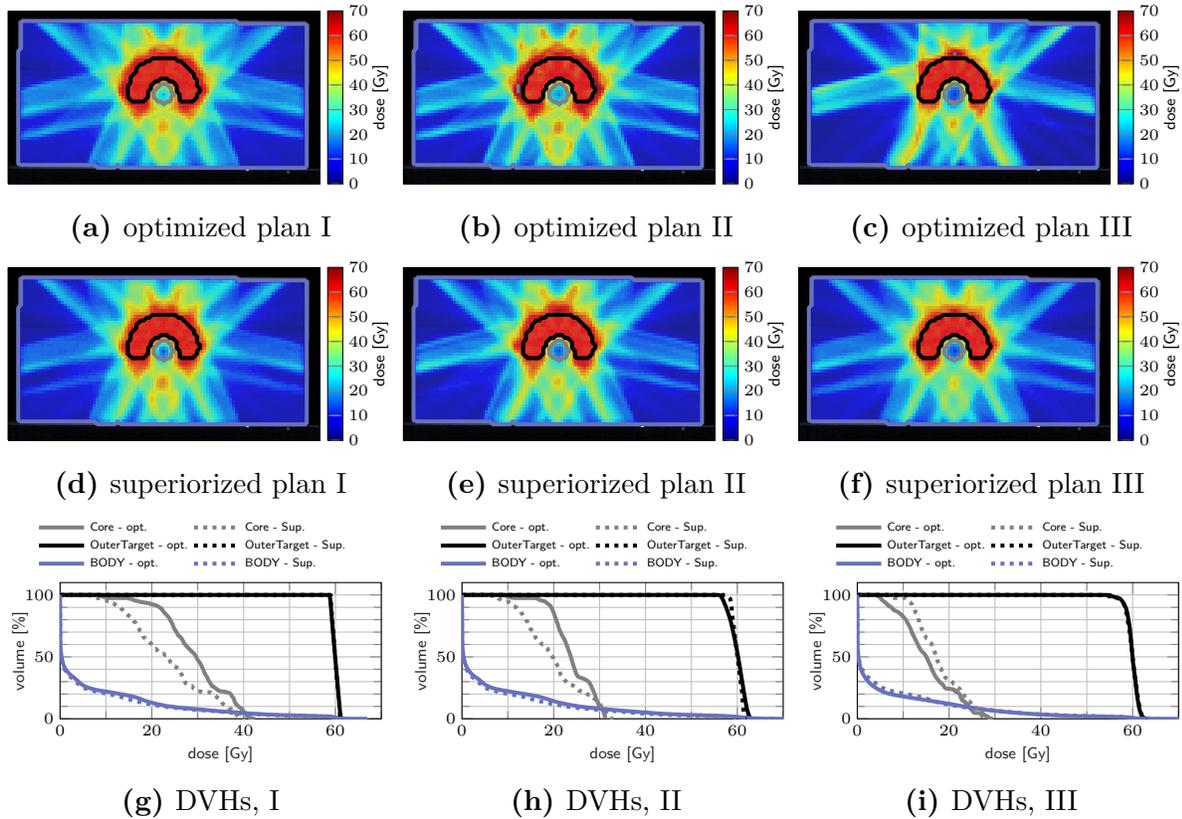

    \begin{subfigure}{0.33\textwidth}
        \centering%
        \tiny\sffamily%
        \setlength{\figurewidth}{0.8\textwidth}%
        \setlength{\figureheight}{0.555\figurewidth}%
        \includetikz{super_script_1_ipoptPlan}%
        \caption{optimized plan I}
        \label{fig:Comparison-of-a:opt1}
    \end{subfigure}%
    \hfill%
    \begin{subfigure}{0.33\textwidth}
        \centering%
        \tiny\sffamily%
        \setlength{\figurewidth}{0.8\textwidth}%
        \setlength{\figureheight}{0.555\figurewidth}%
        \includetikz{super_script_2_ipoptPlan}%
        \caption{optimized plan II}
        \label{fig:Comparison-of-a:opt2}
    \end{subfigure}%
    \hfill%
    \begin{subfigure}{0.33\textwidth}
        \centering%
        \tiny\sffamily%
        \setlength{\figurewidth}{0.8\textwidth}%
        \setlength{\figureheight}{0.555\figurewidth}%
        \includetikz{super_script_3_ipoptPlan}%
        \caption{optimized plan III}
        \label{fig:Comparison-of-a:opt3}
    \end{subfigure}
    
    \vspace{0.5ex}
    
    \begin{subfigure}{0.33\textwidth}
        \centering%
        \tiny\sffamily%
        \setlength{\figurewidth}{0.8\textwidth}%
        \setlength{\figureheight}{0.555\figurewidth}%
        \includetikz{super_script_1_superPlan}%
        \caption{superiorized plan I}  
    \end{subfigure}%
    \hfill%
    \begin{subfigure}{0.33\textwidth}
        \centering%
        \tiny\sffamily%
        \setlength{\figurewidth}{0.8\textwidth}%
        \setlength{\figureheight}{0.555\figurewidth}%
        \includetikz{super_script_2_superPlan}%
        \caption{superiorized plan II}
    \end{subfigure}%
    \hfill%
    \begin{subfigure}{0.33\textwidth}
        \centering%
        \tiny\sffamily%
        \setlength{\figurewidth}{0.8\textwidth}%
        \setlength{\figureheight}{0.555\figurewidth}%
        \includetikz{super_script_3_superPlan}%
        \caption{superiorized plan III}
    \end{subfigure}
    
    \vspace{0.5ex}
    
    \begin{subfigure}{0.33\textwidth}
        \centering%
        \tiny\sffamily%
        \setlength{\figurewidth}{0.825\textwidth}%
        \setlength{\figureheight}{0.35\textwidth}%
        \includetikz{super_script_1_dvhs}%
        \caption{\acp{DVH}, I}
    \end{subfigure}%
    \hfill%
    \begin{subfigure}{0.33\textwidth}
        \centering%
        \tiny\sffamily%
        \setlength{\figurewidth}{0.825\textwidth}%
        \setlength{\figureheight}{0.35\textwidth}%
        \includetikz{super_script_2_dvhs}%
        \caption{\acp{DVH}, II}
    \end{subfigure}%
    \hfill%
    \begin{subfigure}{0.33\textwidth}
        \centering%
        \tiny\sffamily%
        \setlength{\figurewidth}{0.825\textwidth}%
        \setlength{\figureheight}{0.35\textwidth}%
        \includetikz{super_script_3_dvhs}%
        \caption{\acp{DVH}, III}
    \end{subfigure}
    
    \caption{Comparison of treatment plans obtained by superiorization and by constrained minimization. The top row (a)--(c) shows axial dose distribution slices after constrained minimization, the middle row (d)--(f) shows axial dose distribution slices after superiorization. The corresponding \acp{DVH} are shown in the bottom row (g)--(i).}
    \label{fig:Comparison-of-a}
\end{figure}

\begin{figure}[htb]
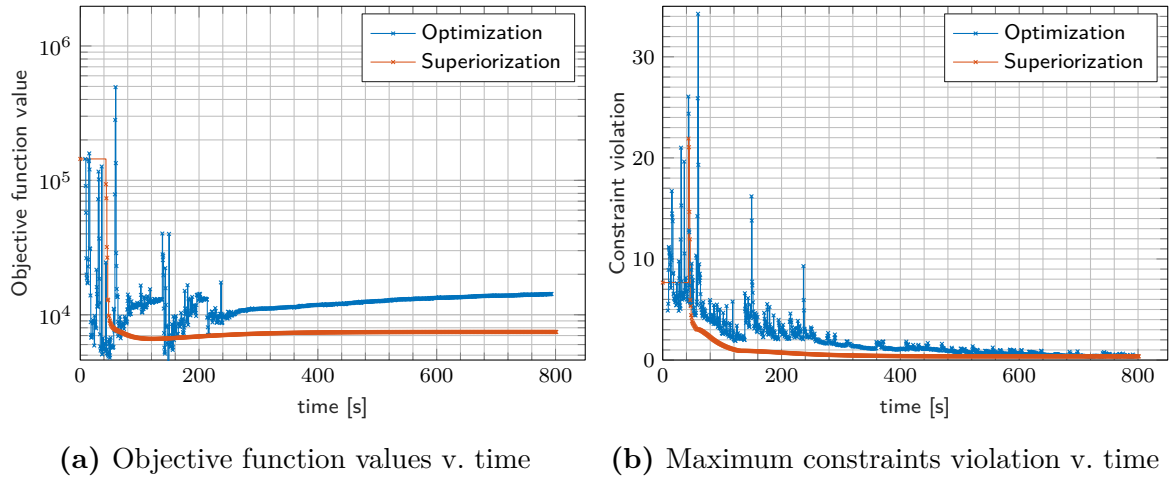

	\begin{subfigure}{0.5\textwidth}
		\centering%
		\scriptsize\sffamily%
		\setlength{\figurewidth}{0.9\textwidth}%
		\setlength{\figureheight}{0.6\textwidth}%
		\includetikz{super_script_1_objvstime}%
		\caption{Objective function values v.\ time}
	\end{subfigure}%
	\hfill%
	\begin{subfigure}{0.5\textwidth}
		\centering%
		\scriptsize\sffamily%
		\setlength{\figurewidth}{0.9\textwidth}%
		\setlength{\figureheight}{0.6\textwidth}%
		\includetikz{super_script_1_constrvstime}%
		\caption{Maximum constraints violation v.\ time}
	\end{subfigure}
\caption{Objective Function values \textbf{(a)} and maximum constraint violation \textbf{(b)} over time for plan I shown in \cref{fig:Comparison-of-a}. \label{fig:Objective-Function-values}. Each cross indicates a full iteration.}
\end{figure}

\Cref{fig:Comparison-of-a} compares dose distributions and
\acp{DVH} after superiorization and after constrained minimization. The respective evolution of the objective function values and the constraint violations (calculated by the infinity norm over all inequality constraint functions, corresponding to the maximum residual) is exemplarily shown in \cref{fig:Objective-Function-values} for plan I.

Comparing plan quality, both plans adhere  to the linear inequality constraints when the problem is feasible (which is the case for plans I \& III) as seen in the \acp{DVH}. In plan I, superiorization appears to reach better \ac{OAR} sparing with reduced mean and maximum dose, while in plan III constrained minimization achieves better \ac{OAR} sparing. For plan II, which poses an infeasible problem, both target coverage and mean \ac{OAR} sparing are improved for superiorization, yet at higher \ac{OAR} maximum dose than obtained through constrained minimization.

The evolution of the objective function and constraint violation for plan I in \cref{fig:Objective-Function-values} exhibits
a \enquote{typical} behavior of superiorization, seeing
a strong decrease in the objective function values within the first iterations,
followed by a slower slight increase as the perturbations' step-sizes diminish. Both approaches were stopped after the maximum number of iterations (\num{1000}) was reached.

Nearly similar constraint violation is achieved by both methods, while constrained minimization resulted in higher objective function values than superiorization, which can be attributed to the difference in \ac{OAR} sparing. For all investigated plans I--III, superiorization showed a much \enquote{smoother} evolution of objective function and constraint violation than observed in the constrained minimization approach.

\subsection{Head-and-neck case}
\label{sec:super_hn}

To prove the usability of superiorization in a conventional planning setting, we applied the \ac{SM} to a head-and-neck case with a wider range of
available objective functions, i.e., including common \ac{DVH}-based objectives.

Coverage of the \acp{PTV} was enforced using voxel inequality constraints.
Again, the results of superiorization were compared to those obtained by solving the constrained minimization problem. All objectives and constraints are given in \cref{tab:prostate_settings}.

\begin{table}[htb]
\centering\caption{Dose inequality constraints, objective functions and penalty weights used for optimization and for superiorization on the head-and-neck case. The functions in the right-hand side column are identified here in \cref{app:objectives}.}
\small%
\begin{tabular}{cccc}
\toprule
VOI & $w_p$ & $c(\boldsymbol{d})$ & $f(\boldsymbol{d})$\tabularnewline
\midrule
PTV70 & 1000 & $\SI{66.5}{\gray} < \boldsymbol{d} < \SI{77}{\gray}$ & $f_{\mathrm{sqdev}}(\boldsymbol{d};\SI{70}{\gray})$\tabularnewline
PTV63 & 1000 &  & $f_{\mathrm{sqdev}}(\boldsymbol{d};\SI{63}{\gray})$\tabularnewline
PTV63 & 1000 &  & $f_{\mathrm{min}\mathrm{DVH}}(\boldsymbol{d};\SI{60}{\gray},\SI{95}{\percent})$\tabularnewline
Spinal Cord PRV & 100 & $\boldsymbol{d} < \SI{50}{\gray}$ & $f_{\mathrm{sqdev}+}(\boldsymbol{d};\SI{15}{\gray})$\tabularnewline
Parotid L \& R & 100 &  & $f_{\mathrm{sqdev}+}(\boldsymbol{d};\SI{10}{\gray})$\tabularnewline
Optic Nerve L \& R & 100 &  & $f_{\mathrm{max}\mathrm{DVH}}(\boldsymbol{d};\SI{50}{\gray},\SI{10}{\percent})$\tabularnewline
Larynx & 300 &  & $f_{\mathrm{sqdev}+}(\boldsymbol{d};\SI{15}{\gray})$\tabularnewline
Chiasm & 100 &  & $f_{\mathrm{max}\mathrm{DVH}}(\boldsymbol{d};\SI{50}{\gray},\SI{10}{\percent})$\tabularnewline
Cerebellum & 100 &  & $f_{\mathrm{sqdev}+}(\boldsymbol{d};\SI{15}{\gray})$\tabularnewline
Brainstem PRV & 100 & $\boldsymbol{d} < \SI{30}{\gray}$ & $f_{\mathrm{sqdev}+}(\boldsymbol{d};\SI{15}{\gray})$\tabularnewline
NT / Body & 100 &  & $f_{\mathrm{mean}}(\boldsymbol{d})$\tabularnewline
\bottomrule
\end{tabular}\label{tab:prostate_settings}
\end{table}

Both solvers use the same stopping criteria for the maximum constraint
violation (smaller than \SI{0.01}{\gray} is acceptable) and objective function
change of value (smaller than \SI{0.1}{\percent} in three consecutive iterations/sweeps).

\begin{figure}[h!]
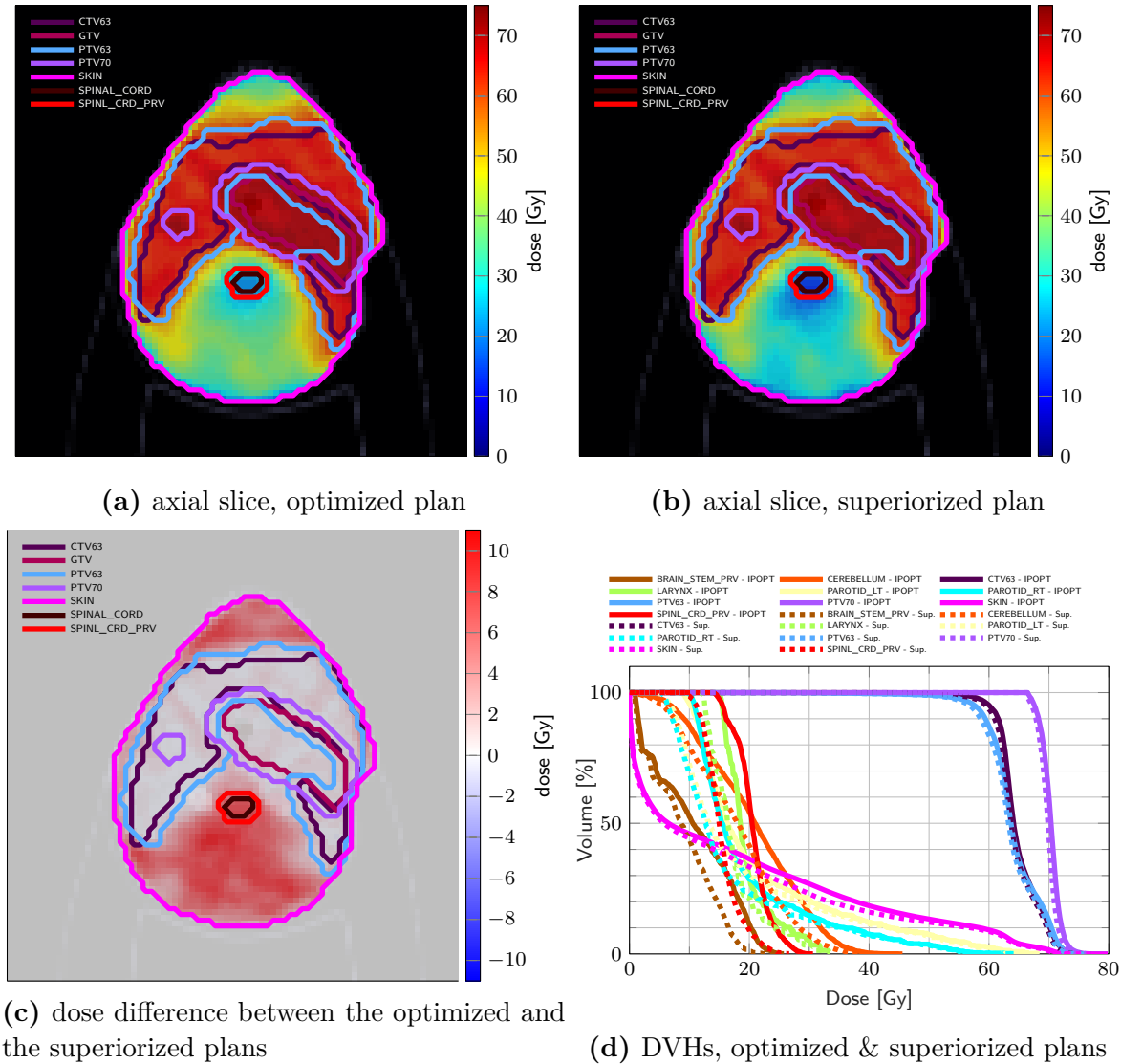

    \begin{subfigure}{0.5\textwidth}
        \centering%
        \scriptsize\sffamily%
        \setlength{\figurewidth}{0.8\textwidth}%
        \setlength{\figureheight}{\figurewidth}%
        \includetikz{super_script_hn_ipoptPlan}%
        \caption{axial slice, optimized plan}     
    \end{subfigure}%
	\hfill%
    \begin{subfigure}{0.5\textwidth}
        \centering%
        \scriptsize\sffamily%
        \setlength{\figurewidth}{0.8\textwidth}%
        \setlength{\figureheight}{\figurewidth}%
        \includetikz{super_script_hn_superPlan}%
        \caption{axial slice, superiorized plan} 
    \end{subfigure}
    
    \vspace{0.5ex}%
    
    \begin{subfigure}[b]{0.5\textwidth}
        \centering%
        \scriptsize\sffamily%
        \setlength{\figurewidth}{0.8\textwidth}%
        \setlength{\figureheight}{\figurewidth}%
        \includetikz{super_script_hn_diffPlan}%
        \caption{dose difference between the optimized and the superiorized plans}
    \end{subfigure}%
    \begin{subfigure}[b]{0.5\textwidth}
        \centering%
        \scriptsize\sffamily%
        \setlength{\figurewidth}{0.85\textwidth}%
        \setlength{\figureheight}{0.51\textwidth}%
        \includetikz{super_script_hn_dvhs}%
        \caption{\acp{DVH}, optimized \& superiorized plans}
    \end{subfigure}
    \caption{Comparison of treatment plans after constrained minimization with IPOPT
and after superiorization (with \ac{AMS} as the basic algorithm) using the tolerances from \cref{tab:prostate_settings}.}
    \label{fig:prostate}
\end{figure}

\begin{figure}[hbt]
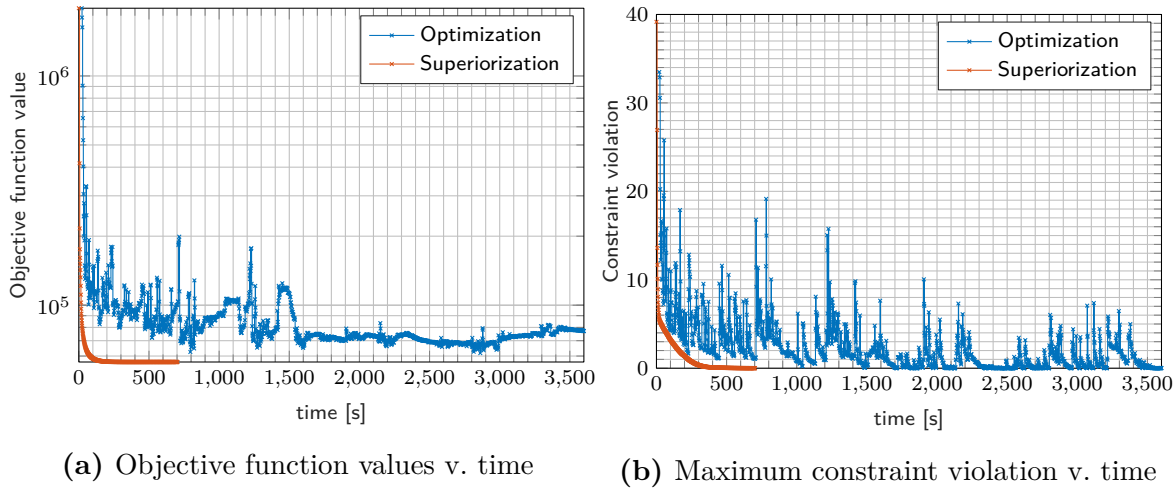

    \begin{subfigure}{0.5\textwidth}
        \centering%
        \scriptsize\sffamily%
        \setlength{\figurewidth}{0.9\textwidth}%
        \setlength{\figureheight}{0.6\textwidth}%
        \includetikz{super_script_hn_objvstime}%
        \caption{Objective function values v.\ time}
    \end{subfigure}%
    \hfill%
    \begin{subfigure}{0.5\textwidth}
        \centering%
        \scriptsize\sffamily%
        \setlength{\figurewidth}{0.9\textwidth}%
        \setlength{\figureheight}{0.6\textwidth}%
        \includetikz{super_script_hn_constrvstime}%
        \caption{Maximum constraint violation v.\ time}
    \end{subfigure}
    
    \caption{Evolution of objective Function values \textbf{(a)} and constraint violation \textbf{(b)} with run-time for the plan shown in  \cref{fig:prostate}.} \label{fig:Objective-Function-values-prostate}
\end{figure}

\Cref{fig:prostate} shows exemplary axial dose slices and the \acp{DVH} for
the plans generated with constraint minimization and with the \ac{SM}.
Quantitative run-time information and evolution of objective function and constraint violation are provided in \cref{fig:Objective-Function-values-prostate}.

Both techniques were able to generate a plan that satisfies the linear
inequalities up to the allowed violation threshold. Considering absolute run-time,
the plan generated with the \ac{SM} satisfied the stopping criteria after \SI{400}{\second}, with constrained minimization failing to converge until the maximum number of iterations was reached. 

\Ac{SM} spent most of the time in the first sweep/iteration, where it focuses on multiple objective function evaluations to generate a large initial decrease (as already observed above). It continuously decreases the objective function values together with decreasing constraints violation, reaching acceptable constraints violation more slowly than the run with constrained minimization. 

However, using the same stopping criteria,
the \ac{SM} reached a solution with a much lower objective function value (approximately one-third of the value achieved by the constrained minimization plan). This is also visible in the dose slices and \ac{DVH}, which show more normal tissue/\ac{OAR} sparing for the \ac{SM} plan. All results are, naturally, only valid for the experiments we performed. Further work, with varying algorithmic parameters, initialization points, and stopping criteria, is necessary to make more general statements.
\section{Discussion}
\label{sec:discussion}

In this work, we applied the novel superiorization method, which solves a system of linear inequalities while reducing a nonlinear objective function, to inverse radiotherapy treatment planning. On a phantom and on a head-and-neck case, we demonstrated that superiorization can produce treatment plans of similar quality to plans generated with
constrained minimization. 

Superiorization showed a smooth convergence behavior for both objective function reduction and constraint violation decrease, including the \enquote{typical} behavior of strong initial objective function reduction with subsequent diminishing objective function reduction -- including potential slight increase -- while proximity to the feasible set within the dose inequality constraints is achieved.

\subsection{The mathematical framework of constrained minimization and of superiorization for treatment planning}

At the heart of the superiorization algorithm lies a feasibility-seeking
algorithm (in this work, the \ac{AMS} relaxation method for linear inequalities). This means that superiorization handles the treatment planning problem as a feasibility-seeking problem for linear inequality dose constraints that should be fulfilled while reducing (not necessarily minimizing) an objective function along the way. 

Constrained optimization algorithms, on the other hand, tackle the same data, i.\,e., constraints and objective function, as a full-fledged optimization problem. With the IPOPT package, for example, inequality constraints become logarithmic barrier functions and are incorporated as a linear combination into the Lagrangian function, whose minimization then enforces the constraints \citep{Wachter2006}. 

When the problem is hardly feasible, finding the right Lagrange multipliers may then dominate the optimization problem in its initial stages. Superiorization with a feasibility-seeking projection algorithm will smoothly reduce the proximity to the constraints, even for infeasible constrained problems, while the perturbations in the objective function reduction phase reduce the objective function value.

Our current implementation is, however, specifically geared for linear constraints. Yet other works on feasibility-seeking have shown
that other relevant constraints, like, e.\,g., \ac{DVH} constraints, can be incorporated into the feasibility-seeking framework, since they can still be interpreted as linear inequalities on a subset (relative volume) of voxels \citep{Penfold2017,Brooke2020,Gadoue2022}.

\subsection{Comparability of run-time, convergence and stopping criteria}

We demonstrated that feasibility-seeking for inverse \ac{IMRT} treatment planning is practically equivalent to the least-squares approach if similar prescriptions are set. However, obtaining the final solution with feasibility-seeking took more time than with unconstrained minimization with our prototype implementation in Matlab. 


Stopping criteria, convergence and run-times are more comparable when considering the constrained minimization vis-à-vis superiorization. Our prototype superiorization algorithm \enquote{converged}
as fast as the used constrained nonlinear minimization algorithm when using the same objective functions and linear inequalities, exhibiting more smooth progress during the iterations. 

Recognizing the limited scope of the experiments presented here, our results about the superiorization method need further work to become well established. For example, the stopping criteria play a substantial role in both optimization and superiorization. Further modification of the respective parameters may lead to earlier or later stopping of either of the algorithms.

Run-time and convergence of a constrained nonlinear optimization algorithm can be influenced by incorporating second derivatives instead of relying on a low-memory approximation to the quasi-Newton approach. In addition, alternative nonlinear minimum/maximum dose constraint implementations are possible. An advantage of the \ac{SM} is that such \enquote{workarounds} are not necessary.

Despite these limitations, we demonstrated that a straightforward superiorization implementation was able to solve the given treatment planning problem arriving at dosimetrically comparable treatment plans.

\subsection{Dosimetric performance}

The treatment plans obtained with constrained minimization and with superiorization show some dosimetric differences. For the three different linearly constrained setups on the TG199 phantom, these differences were most pronounced on the \ac{OAR}, and less pronounced for the target dosage.

In the setups with target dose inequality constraints, superiorization reached better \ac{OAR} sparing. This may be a result of multiple interacting factors: the strong initial objective function decrease in superiorization pulling down the dose in the \ac{OAR}, and potential too early stopping of the constrained minimizer. 

Further, in the infeasible setting with linear inequality constraints on both target and OAR, superiorization has the advantage that the feasibility-seeking algorithm will still smoothly converge to a proximal point. 
 
The improved \ac{OAR} sparing did not occur when only using dose inequality constraints on the \ac{OAR}. However, in this case, the differences in \acp{DVH} of the \ac{OAR} are only substantial below a dose of \SI{20}{\gray} and, thus, of limited significance, since a piece-wise least-squares objective was used that does not contribute to the objective function at dose values below \SI{20}{\gray}.

The head-and-neck case also reproduces the better \ac{OAR} sparing for all evaluated \acp{OAR}, at slightly reduced target coverage for the non-constrained CTV63 and PTV63. Here, the difference in convergence speed was most significant. Through all cases, the superiorization exhibited the smooth evolution of both objective function value and constraint violation, which in turn suggests robustness against changes in the stopping criteria as well. 

These encouraging results show that superiorization can create acceptable
and apparently \enquote{better} treatment plans. Additional work on more cases or planning benchmarks, with varying tuning parameters of both constrained minimization and superiorization approaches is needed to assess the convergence, run-time, and dosimetric quality of the solutions.

\subsection{Outlook}
With the proof-of-concept put forward in this work, there are
many possible directions to further investigate the application
of superiorization algorithms to the radiotherapy inverse treatment planning problem.
From the perspective of a treatment planner, one may focus
on enabling further constraints, e.g., \ac{DVH}-based constraints, that
are often used in treatment planning. 

Some of these constraints are also representable as modified linear inequalities or convex and non-convex sets and, thus, can efficiently be solved using a feasibility-seeking algorithm. Even nonlinear constraints that are based, for example, on normal-tissue complication probability or equivalent uniform dose could be incorporated in the current definition of the superiorization algorithm if the \enquote{basic algorithm} in the feasibility-seeking phase of the \ac{SM} is replaced by any other perturbation resilient projection method that can handle nonlinear constraints. Such algorithms exist in the literature.

Moreover, superiorization might also be extended to use more complex function reduction steps and inherent criteria. For example, a \enquote{true} backtracking line search could be performed, similar to approaches in optimization, since a perturbation resilient \enquote{basic algorithm} might be able to handle much more complex function reduction steps.

Considering these algorithmic and application-focused improvements,
the \ac{SM} should also be rigorously tested on radiotherapy optimization/inverse planning benchmark problems, like the TROTS dataset \citep{Breedveld2017a}, as soon as it is able to handle the respective problem formulations. With this, transferability to other modalities like protons or volumetric modulated arc therapy (VMAT) is also within reach.

\section{Conclusions}
\label{sec:Conclusions}
We introduced superiorization as a novel inverse planning technique, merging feasibility-seeking for linear inequality dose constraints with objective function reduction.
Our initial comparison of superiorization with constrained minimization using linear dose-inequalities suggests possible dosimetric benefits and smoother convergence. Superiorization is thus a valuable addition to the algorithmic inverse treatment planning toolbox.

\section*{Acknowledgments} We thank Mark Bangert for taking part in the first discussion rounds leading up to this work. The work of Y.\ Censor is supported by the ISF-NSFC joint research program grant No. 2874/19.

\appendix
\makeatletter
\renewcommand\@seccntformat[1]{Appendix \csname the#1\endcsname.\hspace{0.5em}}
\makeatother
\section{Objective Functions}
\label{sec:app_objectives}
The nonlinear objective functions used in this work correspond to
those implemented in matRad \citep[Table 1]{Wieser2017}:

\label{app:objectives}
\begin{eqnarray*}
f_{\mathrm{sqdev}}(\boldsymbol{d};\hat{d}) & = & \frac{1}{n}\sum_{i}(d_{i}-\hat{d})^{2}\\
f_{\mathrm{sqdev+}}(\boldsymbol{d};\hat{d}) & = & \frac{1}{n}\sum_{i}\Theta(d_{i}-\hat{d})(d_{i}-\hat{d})^{2}\\
f_{\mathrm{sqdev-}}(\boldsymbol{d};\hat{d}) & = & \frac{1}{n}\sum_{i}\Theta(\hat{d}-d_{i})(d_{i}-\hat{d})^{2}\\
f_{\mathrm{min}\mathrm{DVH}}(\boldsymbol{d};\hat{d},v) & = & \frac{1}{n}\sum_{i}\Theta(\hat{d}-d_{i})\Theta(d_{i}-D_{v}(\boldsymbol{d}))(d_{i}-\hat{d})^{2}\\
f_{\mathrm{max}\mathrm{DVH}}(\boldsymbol{d};\hat{d},v) & = & \frac{1}{n}\sum_{i}\Theta(d_{i}-\hat{d})\Theta(D_{v}(\boldsymbol{d})-d_{i})(d_{i}-\hat{d})^{2}\\
f_{\mathrm{mean}}(\boldsymbol{d}) & = & \frac{1}{n}\sum_{i}d_{i}
\end{eqnarray*}
All objective functions are evaluated on a dose vector $\boldsymbol{d}$ of
length $n$, which corresponds to the number of voxels in each \ac{VOI}. $D_{v}(\boldsymbol{d})$ is the dose at least received by the volume fraction $v$ and thus corresponds to the respective point in the \ac{DVH}. The parameter $\hat{d}$ represents a prescribed/tolerance dose.

\printbibliography

\end{document}